\documentclass[aps,a4paper,twocolumn,floatfix,showpacs,
groupaddress]{revtex4}

\usepackage{bm,amsmath,amssymb,amsfonts}

\usepackage[dvips]{graphicx}
\usepackage[colorlinks=true,citecolor=blue,urlcolor=red,filecolor=blue]{hyperref}
\usepackage{color}
\definecolor{dgreen}{rgb}{0.0,0.5,0.0}
\begin{document}

\title{\textcolor{red}
{Spectral engineering and tunable thermoelectric behavior in a quasiperiodic ladder network}}

\author{Amrita Mukherjee}
\email{amritaphy92@gmail.com}
\affiliation{Department of Physics, University of Kalyani, Kalyani,
West Bengal-741 235, India}
\author{Atanu Nandy}
\email{atanunandy1989@gmail.com}
\affiliation{Department of Physics, Kulti College, Kulti, Paschim Bardhaman,
West Bengal-713 343, India}

\begin{abstract}
Double-stranded quasiperiodic copper mean arrangement has been studied
in respect of their electronic property and thermoelectric signature. The two-arm network
is demonstrated by a tight-binding Hamiltonian. The eigenspectrum of such aperiodic mesh
that does not convey translational invariance, is significantly dependent on the parameters
of the Hamiltonian. It is observed that specific correlation between the parameters obtained 
from the commutation relation between the on-site energy and overlap integral matrices
can eventually modify the spectral nature and generate \textit{absolutely continuous} energy
spectrum. This part is populated by atypical extended states that has a large localization
length substantiated by the flow of the hopping integral under successive real space renormalization
group method steps. This sounds delocalization of single particle energy states in such
non-translationally invariant networks. Further this can be engineered at will by selective 
choice of the relative strengths of the parameters. This precise correlation has a 
crucial impact on the thermoelectric behavior. Anomalous nature of thermoelectric coefficient 
may inspire the experimentalists to frame tunable thermo-devices. Specific correlations
can help us to tune the continuous band and determine the band position at will.
\end{abstract}
\maketitle
\section{Introduction}
\label{intro}
The unique concept of localization of excitation~\cite{anderson,kramer,abrahams} in disordered media 
has recently covered diverse area of physics beyond its original electronic 
proposition. Starting from acoustics~\cite{hu} to Bose Einstein 
condensates~\cite{billy} as well as in the context of 
plasmonic~\cite{christ,ruting}, phononic~\cite{vasseur}, polaronic excitations~\cite{barinov}
 the central issue
of localization is very pertinent. The key 
idea has gained renewed interest due to the evident potential for the realization 
of localization of optical waves in random media. Recent path-breaking experimental 
observations of Photonic localization~\cite{seba1,seba2} have added additional momentum 
in this realm. Not only that the localization of light using path-entangled photons~\cite{gilead} 
and tailoring of partially coherent light~\cite{svozil} have aroused 
extra inspiration from the realistic point of view. 

Now several observations regarding the interesting variation of the classic 
case of the Anderson localization have drawn much attention over the past 
few years in the context of disordered systems.
This was started from the concept of geometrically correlated disorder in the
distribution of on-site energies, the so called \textit{random dimer model} (RDM)~\cite{dunlap}.
 Not only this, the disordered arrangement of overlap integrals~\cite{zhang1}
or quite recently the controlled engineering of absolutely continuous band structure
in quasi one dimensional or quasiperiodic networks~\cite{sil,rudo6,arunava1,arunava2,arunava3,atanu} - all
those have offered significant physics and prospects of designing novel devices in the
era of advanced nanotechnology and lithography technique.
 Suitable correlation between 
the parameters of the Hamiltonian can lead to generate absolutely continuous bands populated by
delocalized Bloch-like 
single particle eigenfunctions. Controlled engineering of band structure by correlated 
manipulation of parameters or even by tuning the external magnetic perturbation sometimes,
may be useful for fabricating some photonic or thermoelectric 
devices. 

In this communication, we address the possible analytical prescription of engineering the absolutely 
continuous bands populated by extended eigenfunctions in case of two-legged 
ladder network which is essentially made of some well-known quasiperiodic geometry. 
This system in fact does not have any translational invariance. The quasiperiodic 
arrangement has a finite extent in the transverse direction but possesses \textit{end-less} 
axial span. Though problem of band tuning we have discussed here is in the context of 
quasiperiodic geometry, it is needless to mention that it is also equally valid 
for any disordered environment. Particularly we have focused on a
deterministic quasiperiodic 
sequence because the exact analytical treatment is possible using the real space 
renormalization group method in the tight-biding formalism.

\begin{figure}[ht]
\centering
\includegraphics[clip,width=8.5cm,angle=0]{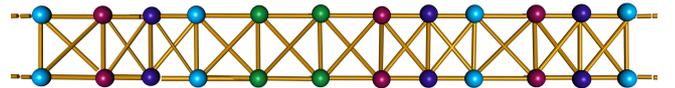}
\caption{(Color online) 
Schematic diagram of two-strand copper-mean ladder network. The different quantum dots are designated 
by different colors such as green dot ($\alpha$ site), red dot ($\beta$ site), purple dot ($\gamma$ site) and
cyan dot ($\delta$ site) respectively. Two types of bonds `long' and `short' bonds 
are assigned with two different 
hopping integrals $t_A$ and $t_B$ respectively.}
\label{cu}
\end{figure}

We concentrate on the ladder network arranged in a 
quasiperiodic Copper-mean (CM)~\cite{aruna} sequence. 
Though any arbitrary number of periodic stacking in the \textit{y}-direction of 
the basic infinitely long two-legged ladder constituting a quasiperioidic mesh 
is in principle possible, we discuss the basic two-arm case only, the multi-channel 
quasiperiodic mesh is just the extended version that contributes nothing additional 
spectral features but only contain rigorous mathematics. We discuss elaborately 
the detailed analysis related to band engineering and its effect on conductivity 
 considering the two-legged CM ladder. This is illustrated in the Fig.~\ref{cu}.

Moreover, the methodology is easily extendable to randomly disordered system of
 arbitrary width.
 In case of random dimer model (RDM), the typical correlated 
 clusters of atomic sites causes local resonances. Whereas here the spectral nature 
 is solely dependent on the competition between the horizontal quasiperiodic 
 ordering and the transverse periodicity (as the system grows vertically). 
 This competition makes the scenario more interesting and challenging. 
 So the deduction of resonance criteria or more specifically speaking, exact 
 numerical correlation between the parameters for the occurrence of Bloch-like 
 extended single particle states thus turns out to be quite non-trivial. The 
 `deterministic' growth pattern of this quasiperiodic sequence helps us 
 to solve the problem in the tight-binding framework.
 Taking the advantage of this we demonstrate the technique of controlled engineering of electronic states in 
quasiperiodic ladder networks and we solidify our analysis by computing the exact numerical correlation, and 
justify it by evaluating the density of states using the Green's function technique. We will obtain the
exact mathematical criteria for generation of \textit{transparent} band and substantiate our discussion by
 computing the transport characteristics.
 
 The strength of our analytical attempt lies in the fact that we can obtain
 an \textit{exact} mapping of a two or multi-arm quasiperiodic ladder network into
 a set of completely `isolated' linear chains that describe the quantum mechanics
 of a class of \textit{pseudoparticles}. Such an exact mapping has already been
 demonstrated in ladder-like geometries~\cite{sil,rudo6,amrita} modelling  
 a DNA-like double chain or a quasi-two dimensional 
 mesh with correlated disorder in the context of de-localization of single particle
 eigenstates.

In the next part,
we study the thermoelectric transport property of these aperiodic ladder geometries in 
the tight-binding formalism. Suitable parametric correlation induced spectral
nature can make the system transparent to the incoming electron for selective zone
of Fermi energy and this selection can be controlled at will. In this work, we try to
explore this correlation dependent transmission profile for studying the thermoelectric
behavior of the system arranged in a quasiperiodic fashion.

The study of charge transport in molecules has potential applications in molecular 
electronics~\cite{aviram} and energy-conversion devices~\cite{ditt,yu} such
as electron transfer, shot noise, heat transport, negative differential resistance
and gate controlled effects. Recently, topics on thermorelated transport such as 
local heating and thermal transport have emerged as new sub-field in molecular electronics.
One of the significant aspect in the molecular tunnel junction is thermoelectricity.
The Seebeck coefficient which is basically the slope of the transmission function in the vicinity 
of Fermi energy level, can carry more significant information than simple current-voltage
characteristics. The study of thermoelectricity has its main importance in designing
some thermorelated electronic and nano scale energy conversion devices.
The experimental study of thermoelectric voltage over guanine molecules adsorbed 
on a graphite substrate reported by Poler et. al.~\cite{poler} has opened
novel perspectives in the quest for thermoelectric devices based on molecular
electronics engineering.
From the theoretical point of view, the extreme response of thermopower to finer details in
the electronic structure allows us to extract considerable information regarding the relative
position of Fermi level compared to molecular levels and also from the practical perspective, in
order to optimize the thermoelectric figure of merit of a given molecular arrangement.
Inspired by lots of experimental measurements of thermopower and comparison with
the theoretical results for some organic molecular systems~\cite{macia,supriyo,gao}
here we try to figure out the flavor of correlation dependent tunable thermoelectric behavior for
this model network within the tight-binding prescription.
We find interesting results. For a multi-arm copper-mean mesh, it is possible to give
an \textit{exact} analytical correlation between the parameters of the Hamiltonian. For 
two-arm case, there can be two such possibilities of different choices of correlation.
For odd-strand network, there will be $N-1$ different choices of correlations, where $N$ is
the number of strand. For all such cases, the overall spectrum loses its quasiperiodic 
identity. The inter-strand connection also plays a significant role in this context. 
The transport behavior is also tested by virtue of usual technique. This gives a strong support
to the
spectral nature.
In the second part of our work, we find the correlation induced thermoelectric property.
It shows an anomalous behavior with energy. This correlation dependence could make the 
system more appreciating in the platform of device technology.

In what follows we describe our findings. In the Section 2, we demonstrate the Hamiltonian and the basic methodology of generating and engineering the continuous resonant band in a quasiperiodic mesh. Section 3 deals with the evaluation of density of states profile and the subsequent relevant discussion regarding the controlled band tuning. Then in Section 4, the technique of transport calculations is presented to justify our analytical treatment. Finally, we draw our conclusions.

\section{Model and Hamiltonian}
\label{cop}
First we will start our discussion with the prototype geometry comprising of two-channel copper-mean ladder 
network as cited in the Fig.~\ref{cu}. 
Ordinary copper-mean chain has two types hopping integral parameters, 
in the tight-binding description, viz., `long' ($A$) and `short' ($B$) ``bonds” and the network grows following 
the chronological sequence: $A \rightarrow ABB$, $B \rightarrow A$. Accordingly,
the successive generations are $G_1 = A$, $G_2 = ABB$, $G_3 = ABBAA$, $G_4 = ABBAAABBABB$ and so on.
Following this quasi-periodic sequence
one can in principle, periodically couple an enormous number of such infinitely extended (along $X$-axis) 
copper-mean arrays in the transverse direction or even without periodicity. We discuss here the $N$-legged 
copper-mean mesh which has periodicity along transverse axis.

Any type of this multi-strand quasi-periodic ladder systems can be described in the standard tight-binding 
formalism by the following Hamiltonian in the Wannier basis for the spinless non-interacting electrons, viz.,

\begin{widetext}
\begin{equation}
H  = \sum_{n,k} \epsilon_{n}^{k} {{c_{n}}^{k}}^{\dagger} {c_{n}}^{k}
+\sum_{\langle nm \rangle,k} t_{nm}^{k} {c_{n}}^{{k}^{\dagger}} 
{c_{m}}^{k} 
+ \sum_{n,\langle kl \rangle} \Gamma_n {c_{n}}^{{k}^{\dagger}} {c_{n}}^{l} + 
\sum_{\langle nm \rangle, k} {\chi_{nm}}^{k,k+1} {c_{n}}^{{k}^{\dagger}} 
{c_{m}}^{k+1} + 
\sum_{\langle nm \rangle, k} {\chi_{nm}}^{k,k-1} {c_{n}}^{{k}^{\dagger}} {c_{m}}^{k-1}
\end{equation}
\label{hamilton}
\end{widetext}

The first part of the Hamiltonian signifies the potential contribution 
while the other terms carry the kinetic contribution.
The pairs of indices ($n$, $m$) are associated with nearest neighbor atomic sites on any particular 
strand, while $k$ and $l$ index represent different strands in the mesh. We can distinguish between the 
different types of quantum dots present in the system. There are basically four kinds of atomic sites in 
each strand depending on the local environment, which are $\alpha$ (green colored spheres) sites, $\beta$ 
(red colored spheres) sites, $\gamma$ (purple colored spheres) sites and $\delta$ (cyan colored spheres) 
kind of sites respectively flanked by pairs of bonds $A-A$, $A-B$, $B-A$ and $B-B$. 
The on-site potentials of these kind of sites are designated as $\epsilon_{\alpha}$, 
$\epsilon_{\beta}$, $\epsilon_{\gamma}$ and $\epsilon_{\delta}$ respectively.
The nearest neighbor hopping integrals in any strand  
are assigned values $t_A$, $t_B$, across the $A$ and the $B$ bonds respectively.
We additionally introduce second neighbor connection
between pairs of strands 
across the diagonals in the 
bigger and the smaller rectangular plaquettes as shown, and 
denote them by ${\chi_{nm}}^{k,k\pm 1} = \chi_A$ or  
$\chi_B$ 
respectively according to the geometry. 
The inter-strand tunnel hopping, connecting the 
$i$-th site in the $k$-th strand with the $n$-th site in the $l$-th 
strand ($l = k \pm 1$) is  
$\Gamma_{n}=\Gamma_\alpha$, $\Gamma_\beta$, $\Gamma_\gamma$ or $\Gamma_\delta$ depending on whether 
it connects $\alpha$, $\beta$, $\gamma$ or $\delta$ sites of the neighboring 
strands along the transverse axis.
The provision of tuning the inter-strand hopping parameters or the second neighbor
hopping integrals makes the model more realistic implying that one
can, in principle, talk about the quasiperiodically \textit{distorted} ladder
networks or double stranded DNA-like models~\cite{sou1,sou2} as well.
 This definitely brings a flavor of geometrical 
disorder and hence we can discuss its effect on the spectral landscape
of such systems within the same formalism. Also, second motivation of this work is to study the
tunable thermo-electric transport in the next part as a result of such quasiperiodically modulated geometry. 

By virtue of the generalized difference equation (which is the 
discretized version of the Schr\"{o}dinger's equation) 
given by
\begin{equation}
(E-\epsilon_{n}) \psi_{n} = \sum_{m} t_{nm} \psi_m
\label{diff1}
\end{equation}
we can easily obtain,
for any such $N$-strand quasiperiodic mesh, $3N$ difference equations.
Therefore, for copper-mean ladder, there are 
$N$-equations corresponding to each vertical rung with $\alpha$, $\beta$, $\gamma$ or $\delta$ 
sites residing on it. To avoid complicacy regarding generalized form of such
difference equations, we 
focus on a two-strand only. It is needles to say that this formalism can be extrapolated 
for any $N$-legged quasi-periodic ladder. 
Moreover, this is enough to bring out the central spirit of the calculations, and a  
generalization to the case of arbitrary $N$ is quite trivial. 

For a two-strand network, the difference equations for an $\alpha$-rung read, 
\begin{widetext}
\begin{eqnarray}
(E - \epsilon_\alpha) \psi_{i,2} & = & t_L \psi_{i+1,2} + t_L \psi_{i-1,2} 
\chi_L \psi_{i+1,1} + \chi_L \psi_{i-1,1} + 
\Gamma_\alpha \psi_{i,1} \nonumber \\
(E - \epsilon_\alpha) \psi_{i,1} & = & t_L \psi_{i+1,1} + t_L \psi_{i-1,1} 
+ \chi_L \psi_{i+1,2} + \chi_L \psi_{i-1,2} + \Gamma_\alpha \psi_{i,2} \nonumber \\
\label{diffalpfa}
\end{eqnarray}
\end{widetext}.
We can similarly write the difference equations for the other rungs considering the local
atmosphere of the corresponding sites.
All these difference equations corresponding to each rung can be easily recast in a combined matrix form viz.,
\begin{equation}
(E \bm{I} - \bm{\epsilon}) \bm{\psi_{n}} = \bm{t}_{n+1} \bm{\psi_{n+1}} + \bm{t}_{n-1} \bm{\psi_{n-1}}
\label{matrix}
\end{equation}
Here, 
\begin{equation}
\bm{I}=
\left( \begin{array}{cccc}
1 & 0 \\
0 & 1 \\

\end{array}
\right)
\label{mat1}
\end{equation}
\begin{equation}
\bm{\epsilon}=
\left( \begin{array}{cccc}
\epsilon_{n,2} & \Gamma_{21} \\
\Gamma_{12} & \epsilon_{n,1} 
\end{array}
\right)
\label{mat2}
\end{equation}
\begin{equation}
\bm{\psi_n}=
\left ( \begin{array}{c}
\psi_{n,2} \\
\psi_{n,1} 
\end{array} \right )
\label{mat3}
\end{equation}
\begin{equation}
\bm{t_{n \pm 1}}=
\left( \begin{array}{cccc}
t_{i,i \pm 1}^{2} & \chi_{i,i \pm 1}^{21} \\
\chi_{i,i \pm 1}^{12} & t_{i,i \pm 1}^{1} 
\end{array}
\right )
\label{mat4}
\end{equation}
\begin{equation}
\bm{\psi_{n \pm 1}}=
\left ( \begin{array}{c}
\psi_{n \pm 1,2} \\
\psi_{n \pm 1,1}
\end{array} \right )
\label{mat5}
\end{equation}

It is to be noted that the `potential matrix' (comprising of the on-site potential and the
inter-strand coupling) and the `hopping matrix' (containing nearest neighbor and second neighbor
connections) commute with each other, and
hence can be simultaneously diagonalized by a similarity transform.
Eq.~\eqref{matrix} can then be easily decoupled, in a new basis defined
by the following relation
\begin{equation}
\left ( \begin{array}{c}
\phi_{2} \\
\phi_{1}
\end{array} \right )
=\bm {M}^{-1} 
\left ( \begin{array}{c}
\psi_{2} \\
\psi_{1}
\end{array} \right )
\label{basischange}
\end{equation}
The matrix $\bm M$ diagonalizes both the `potential'
and the `hopping' matrices. We can make therefore a uniform change of basis 
and in the new basis,
we are left with four independent linear difference equations for each of the rungs.
Each of these equations represents a copper-mean chain describing a
kind of \textit{pseudoparticles} with states that are
linear combinations of the old Wannier orbitals.
The decoupled, independent, linear equations are:
\begin{align}
& \left[ E - (\epsilon_\alpha - \Gamma_\alpha) \right] \phi_{i,2} =
(t_L - \chi_L) \phi_{i+1,2} + \nonumber \\
& \qquad \qquad \qquad \qquad \qquad \qquad (t_L - \chi_L) \phi_{i-1,2} \nonumber \\
& \left[ E - (\epsilon_\beta - \Gamma_\beta) \right] \phi_{i,2} =
(t_S - \chi_S) \phi_{i+1,2} + \nonumber \\
& \qquad \qquad \qquad \qquad \qquad \qquad (t_L - \chi_L) \phi_{i-1,2} \nonumber \\
& \left[ E - (\epsilon_\gamma - \Gamma_\gamma) \right] \phi_{i,2} =
(t_L - \chi_L) \phi_{i+1,2} + \nonumber \\
& \qquad \qquad \qquad \qquad \qquad \qquad (t_S - \chi_S) \phi_{i-1,2} \nonumber \\
& \left[ E - (\epsilon_\delta - \Gamma_\delta) \right] \phi_{i,2} =
(t_S - \chi_S) \phi_{i+1,2} + \nonumber \\
& \qquad \qquad \qquad \qquad \qquad \qquad (t_S - \chi_S) \phi_{i-1,2}
\label{decouple1}
\end{align}
\begin{align}
& \left[ E - (\epsilon_\alpha + \Gamma_\alpha) \right] \phi_{i,1} =
(t_L + \chi_L) \phi_{i+1,1} + \nonumber \\
& \qquad \qquad \qquad \qquad \qquad \qquad (t_L + \chi_L) \phi_{i-1,1} \nonumber \\
&\left[ E - (\epsilon_\beta + \Gamma_\beta) \right] \phi_{i,1} =  
(t_S + \chi_S) \phi_{i+1,1} + \nonumber \\
& \qquad \qquad \qquad \qquad \qquad \qquad (t_L + \chi_L) \phi_{i-1,1} \nonumber \\
&\left[ E - (\epsilon_\gamma + \Gamma_\gamma) \right] \phi_{i,1} = 
(t_L + \chi_L) \phi_{i+1,1} + \nonumber \\ 
& \qquad \qquad \qquad \qquad \qquad \qquad (t_S + \chi_S) \phi_{i-1,1} \nonumber \\
&\left[ E - (\epsilon_\delta + \Gamma_\delta) \right] \phi_{i,1} = 
(t_S + \chi_S) \phi_{i+1,1} + \nonumber \\ 
& \qquad \qquad \qquad \qquad \qquad \qquad (t_S + \chi_S) \phi_{i-1,1}
\label{decouple3}
\end{align}
Each of the equations Eq.~\eqref{decouple1} $-$ Eq.~\eqref{decouple3} 
yields two independent copper-mean chains with the corresponding effective 
on-site potentials [$(\epsilon_\alpha \pm \Gamma_\alpha)$, 
$(\epsilon_\beta \pm \Gamma_\beta)$, 
$(\epsilon_\gamma \pm \Gamma_\gamma)$, $(\epsilon_\delta \pm \Gamma_\delta)$].
Also, the effective hopping integral parameters are $t_L \pm \chi_L$, 
$t_S \pm \chi_S$.
Further note that when we consider individual set of the equations, each
set gives rise to typical spectrum of copper mean chain containing a 
broad continua populated by extended eigenfunctions (or eigenfunctions with large
localization lengths). Thus, it is needless to say that
the
actual spectra of the CM-ladder network is just the convolution of the individual spectrum.
Now we should make comment on the nature of the eigenstates for this multi-legged 
CM ladder network before ending this section. This can be made possible if we look 
at the decoupled equations. From the basis transformation formalism, one can 
easily say that the wavefunctions $\phi_i$'s in the changed basis are the linear superposition
of the original amplitudes $\psi_i$'s. Therefore, localization
character of any $\phi_i$ will prevail only if all the contributing
eigenfunctions carry the same character. On the contrary, if by means of some
correlations between the parameters, one can make any of the ``channel'' transparent
to the incoming wave-train, then it will render the entire linear combination
$\phi_i = \sum_{j} \psi_j$ conducting.

\subsection{Spectral Information by RSRG method}
\label{density}
To obtain the spectral landscape of this
multi-strand quasi-periodic ladder model,
we use the standard \textit{real space renormalization
group method} using the potential matrix and hopping matrix respectively
given by the Eq.~\eqref{mat2} and Eq.~\eqref{mat4}.
In this scheme, we decimate out an appropriate subset of the vertices in terms
of the amplitudes of the surviving nodes. Thus for the CM ladder, the
decimation procedure is implemented via the backward \textit{folding}, viz.,
$ABB \rightarrow A^{'}$ and $A \rightarrow B^{'}$.
It is simple to work out the RSRG recursion relations containing the
renormalized potential and hopping and are given by,
\begin{widetext}
\begin{eqnarray}
\bm{\epsilon}_{\alpha,n+1} & = & \bm{\epsilon}_{\gamma,n} + 
\bm{t}_{S,n}^{T} (E.\bm{I} - \bm{\delta}_{1,n})^{-1} \bm{t}_{S,n} +
\bm{t}_{L,n} (E.\bm{I} - \bm{\delta}_{2,n})^{-1} \bm{t}_{L,n}^{T} \nonumber \\
\bm{\epsilon}_{\beta,n+1} & = & \bm{\epsilon}_{\gamma,n} + 
\bm{t}_{S,n}^{T} (E.\bm{I} - \bm{\delta}_{1,n})^{-1} \bm{t}_{S,n} \nonumber \\
\bm{\epsilon}_{\gamma,n+1} & = & \bm{\epsilon}_{\alpha,n} +      
\bm{t}_{L,n} (E.\bm{I} - \bm{\delta}_{2,n})^{-1} \bm{t}_{L,n}^{T} \nonumber \\
\bm{\epsilon}_{\delta,n+1} & = & \bm{\epsilon}_{\alpha,n}         \nonumber \\
\bm{t}_{L,n+1} & = & \bm{t}_{L,n} (E.\bm{I} - \bm{\epsilon}_{\beta,n})^{-1} 
\bm{t}_{S,n} (E.\bm{I} - \bm{\delta}_{1,n})^{-1} \bm{t}_{S,n} \nonumber \\
\bm{t}_{S,n+1} & = & \bm{t}_{L,n} \nonumber \\
\bm{\delta}_{1,n} & = & \bm{\epsilon}_{\delta,n} + 
\bm{t}_{S,n}^{T} (E.\bm{I} - \bm{\epsilon}_{\beta,n})^{-1} \bm{t}_{S,n} \nonumber \\
\bm{\delta}_{2,n} & = & \bm{\epsilon}_{\beta,n} + 
\bm{t}_{S,n} (E.\bm{I} - \bm{\epsilon}_{\delta,n})^{-1} \bm{t}_{S,n}^{T}
\label{curecursion}
\end{eqnarray}
\end{widetext}
For the numerical evaluation of the density of states, we add a small imaginary
part to the energy. Following the above recursion relations the hopping integrals
flow to zero after certain number of iterations and potential part will reach to its 
fixed point value. We then calculate the local green's function
$\bm {G} = (E.\bm {I} - \bm {\epsilon_i}^{*})^{-1}$ where the subscript $i$ 
may be different corresponding to the different type of nodes. The `asterix' mark
denotes the fixed point value of the on-site term. Finally, the diagonal term
of the Green's function matrix will give the local spectral information of any particular
site.

\section{Band Engineering}
\label{bandeng}
The advantage of the decoupling scheme lies in the fact that one can have the
\begin{figure}[ht]
\centering
\includegraphics[clip,width=8.5cm,angle=0]{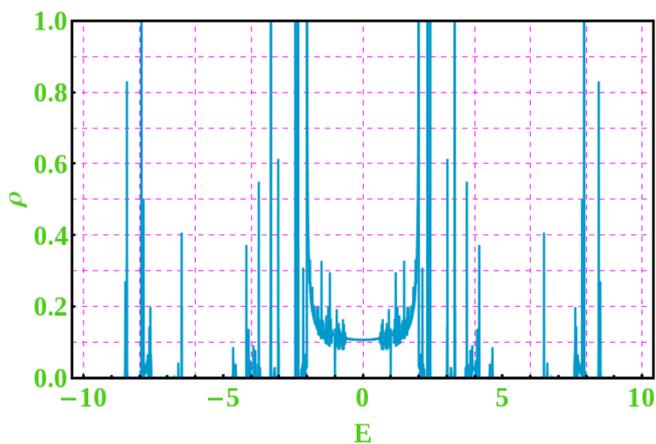}
\caption{(Color online) 
(a) The spectral landscape of Density of States with Energy. 
We have set 
$\epsilon_i=0$, $t_L=1$ and, $\chi_L=2, \chi_S=3$. Also,
all the vertical couplings $\Gamma_{\alpha}$, $\Gamma_{\beta}$ and
$\Gamma_{\gamma}$ are set equal to $\Gamma$ for numerical calculation.}
\label{cu2dos}
\end{figure}
liberty to engineer the extent of absolutely continuous energy bands even in this kind of system
where there is no question of translational invariance. Such type of correlation does not
bother about the individual values of the parameters of the Hamiltonian. So, taking in terms of
some realistic DNA kind of model, one can choose the relevant values of the parameters taken from the
simulation technique. We justify the above claim by pointing out to the fact that, for example,
in equation~\eqref{decouple1}, if we set all $\epsilon_i - \Gamma_i$ same and equal to 
constant for all $i$ and $t_L - \chi_L = t_S - \chi_S$, then the set of 
equation~\eqref{decouple1} represents a perfectly ordered
system with the corresponding band edge extending from
$E = (\epsilon_\alpha - \Gamma_\alpha) - 2 (t_L - \chi_L)$ to
$E = (\epsilon_\alpha - \Gamma_\alpha) + 2 (t_L - \chi_L)$.
Note that this continuous bands will be populated by \textit{extended}, Bloch-like
pseudo-particle eigenstates. 

It is to be that the correlation made for band engineering does not impose
any restriction on the individual values of the on-site potential $\epsilon_i$ and 
inter-strand connection $\Gamma_i$. This only needs that the differences $\epsilon_i \pm \Gamma_i$
and $t_{ij}^{k} - \chi_{ij}^{k,k \pm 1}$ have to remain unchanged. In an exactly
similar fashion, if we look at the other set of decoupled difference equation for this
quasiperiodic mesh, parallel observation can be obtained, where the correlations
needed for the generation of resonant conducting band are $(\epsilon_\alpha + \Gamma_\alpha)
= (\epsilon_\beta + \Gamma_\beta) = (\epsilon_\gamma + \Gamma_\gamma) = (\epsilon_\delta + \Gamma_\delta)$
and $t_L + \chi_L = t_S + \chi_S$.

In Fig.~\ref{cu2dos}, we show the density of states profile of a two strand copper-mean ladder
with all on-site energy $\epsilon_i=0$ and following the correlations mentioned above. Here, we
have set, without any loss of generality, $\Gamma_\alpha = \Gamma_\beta = \Gamma_\gamma = \Gamma_\delta = 0$,
just to make the center of the spectrum at $E=0$, and the inter-strand connection happens through
the diagonal hopping integral term only. Now it is to be mentioned that any one of those two decoupled
difference equations will at a time contribute to generate an absolutely continuous band populated
by extended Bloch-like states only. The other equation will then populate the spectrum with critical
eigenstates, typical characteristics of the quasiperiodic arrangement. The convolution of those two different
eigenstates obtained from two decoupled equations under parametric correlation will demonstrate the 
complete spectral information. Therefore, if some of the critical states happen to be occupying
part of the spectrum that falls within the central continuous zone, then they will loss their
critical identity and become a member of extended family. This is actually what happens in the 
Fig.~\ref{cu2dos}. Here we incorporate the correlation between the nearest neighbor hopping
integrals and next nearest connection as $t_L + \chi_L = t_S + \chi_S$ with $t_L=1$, $t_S=2$, 
$\chi_L = 2$, $\chi_S = 3$. As soon as we set the correlation, one difference equation
represents a perfectly periodic chain of identical atomic sites that contributes an
absolutely continuous band populated by extended eigenfunctions having extent between 
$-2 \leq E \leq 2$. The other difference equation will obviously give 
the spectrum that carries the flavor of quasiperiodic copper-mean geometry. The full spectrum,
as obtained from RSRG recursion relation and the Green's functions reproduce the absolutely
continuum exactly over the energy regime, as extracted from decoupled equation. (Eq.~\eqref{decouple1}).

The resonant character of those states belonging to the central continuum can be easily
justified following the usual method of checking the flow of hopping integrals under 
successive RSRG iterations. We have carefully scanned the continuous zone over arbitrary small
energy intervals. For any arbitrary energy picked up from the continuum, the hopping
integral retain oscillatory nature without showing any signature of convergence to zero for
an arbitrary large number of loops. This brings a clear indication that the corresponding 
Wannier orbitals have finite non-zero overlap over arbitrarily large distances - basic characteristic
feature of the extended eigenfunction.

In a similar manner it is very easy to perform the same checking of the flow of hopping parameter under
 successive iteration steps for the single particle states situated at the flank of the eigenspectrum. 
 The flow will converge to zero after a moderate number (small compared to that of central transparent states) 
 of RSRG steps. This immediately suggests that the corresponding eigenfunctions at least have very large
  localization lengths (if not considered to be extended). At this point we should mention that we have used 
  the above decoupled equations only to extract the region of the central (in this case)
continuous subband. The profile of density of eigenstates presented here is obtained by using the RSRG 
method on the full two-strand quasiperiodic mesh.
\section{Transport Characteristics}
\label{transmission}
\begin{figure}[ht]
\centering
\includegraphics[clip,width=8.5cm,angle=0]{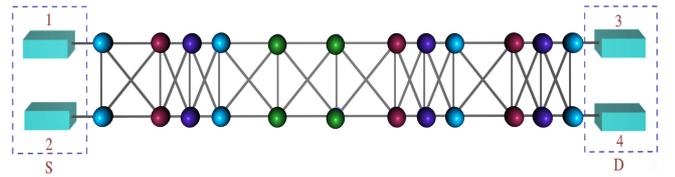}
\caption{(Color online) 
A finite size coppermean ladder network is sandwiched between semi-infinite leads.
The different quantum dots are designated 
by different colors such as green dot ($\alpha$ site), red dot ($\beta$ site), purple dot ($\gamma$ site) and
cyan dot ($\delta$ site) respectively. Two types of bonds `long' and `short' bonds 
are assigned with two different 
hopping integrals $t_A$ and $t_B$ respectively.}
\label{cufinal}
\end{figure}

To corroborate our findings we will now do parametric calculation to elucidate the transport characteristics of the quasiperiodic coppermean ladder like systems following the usual method of multichannel transport. 
For this purpose, we need to focus our attention on the systems where a finite size two strand coppermean ladder network is connected with a pair of semi-infinite leads as shown in the Fig.~\ref{cufinal}. To evaluate the transmittance, here we now adopt the Green's
 function approach. Hence we need to define the Green's function for the composite systems (lead-system-lead) which is given by,
\begin{equation}
G= (E-H)^{-1}
\label{transeq1}
\end{equation}
where $E=\epsilon+i\eta$ with $\eta$ being the arbitrarily small number which can be set as zero in the limiting approximation. $\epsilon$ is the energy of the incoming electron. H is the Hamiltonian of the entire system including the semi-infinite leads also. So the above Green's function equation deals with the inversion of the matrix of infinite dimension that corresponds to a system consisting of the finite size quasiperiodic mesh and semi-infinite leads. Therefore, if we write the Hamiltonian explicitly for the individual subsystem then we have,
\begin{equation}
H=H_M+\sum_{m=1}^N(H_m+H_{mM}+H_{mM}^\dagger)
\label{transeq2}
\end{equation}
where $H_M$ and $H_m$ are the Hamiltonians respectively for the original finite size 
ladder geometry and lead-m-N corresponds to the number of leads to which the system gets connected. $H_{mM}$ denotes the coupling matrix. Further note that all the leads designated with the respective tight-binding parameters have equal footing. also within the tight-binding framework the system is demonstrated by the above Hamiltonian as written in Eq.~\eqref{transeq2}.
 Following the partition of the Hamiltonian, we can easily write the effective Green's function for the system introducing the Lowdin's partitioning technique~\cite{low1,low2} as,
\begin{equation}
G_M=(E-H_M-\sum_{m=1}^N\Sigma_m)^{-1}
\label{transeq3}
\end{equation}
where $\Sigma_m$ offers the self-energy contribution because of the presence of the coupling of the system to the lead. The explicit expression for the self energy corresponding to particular $m^{th}$ lead is given by, 
\begin{equation}
\Sigma_m=H_{mM}^{\dagger}G_mH_{mM}
\label{transeq4}
\end{equation}
Here $G_m=(E-H_m)^{-1}$ is the Green's function of $m^{th}$ lead. It is needless to mention that the self energy
 term carries the signature of the entire information regarding coupling between the system and lead. 
Using the Eq.~\eqref{transeq4}, the coupling function $\Gamma_m$ can be easily obtained from the following
 equation~\cite{su1,su2}, 
\begin{equation}
\Gamma_m(E)=i[\Sigma_m^r(E)-\Sigma_m^a(E)]
\label{transeq5}
\end{equation} 
where $\Sigma_m^{a(r)}(E)$ signifies the advanced (retarded) self energy terms respectively. It is obvious that they are Hermitian conjugate to each other.
Thereafter, we may rewrite the above Eq.~\eqref{transeq5} as,
\begin{equation}
\Gamma_m=-2Im(\Sigma_m^r)
\label{transeq6}
\end{equation}
Now the transmission co-efficient can be expressed in terms of the effective system and system-to-lead coupling as,
\begin{equation}
T_{mn}=Tr[\Gamma_mG_M^r\Gamma_nG_M^a]
\label{transeq7}
\end{equation}
 Here $\Gamma_m$ and $\Gamma_n$ denote the connection of the system to the $m^{th}$ and $n^{th}$ lead respectively and $G-M^r$ and $G_M^a$ are the retarded and advanced Green's function of the system respectively.
 As the coupling matrix $H_{mM}$ is non-zero only for the adjacent points, the expression for the 
 transmission probability becomes~\cite{kemp}
 \begin{equation}
T_{mn}=4 \Delta_{m} \Delta_{n} |G_{M}|^2
\label{transeq8}
\end{equation}
 
 The transmission characteristics for this two-strand copper mean ladder network is represented in the
\begin{figure}[ht]
\centering
\includegraphics[clip,width=8.5cm,angle=0]{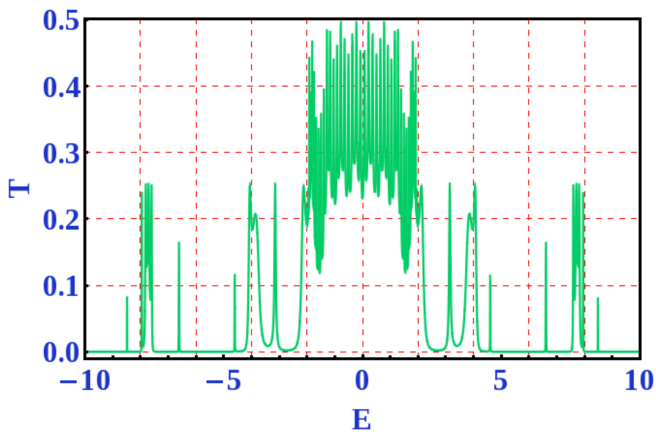}
\caption{(Color online) 
(a) The transmission characteristics of a coppermean ladder 
with 43 bonds in each arm as a function of energy under the specific correlation. 
We have set 
$\epsilon_i=0$, $t_L=1$ and, $\chi_L=2, \chi_S=3$. Also,
all the vertical couplings $\Gamma_{\alpha}$, $\Gamma_{\beta}$ and
$\Gamma_{\gamma}$ are set equal to $\Gamma$ for numerical calculation.}
\label{cu2trans}
\end{figure}
  Fig.~\ref{cu2trans} using the above analytical
expression~\ref{transeq8}. From the graphical plot, it is clear that the system shows resonant behaviour to the incoming electron for the central regime of energy ($ -2 \le E \le 2 $) where it shows a moderately high transport. Outside this central conducting zone, there exists few discrete set of energies at the flank of the spectrum for which it is showing relatively low transport. The conducting nature of the system basically reflects the signal of quasiperiodic geometry present. The central resonating region is happening due to the arrangements of `short' bonds.
 
\section{Thermoelectric behavior of copper mean ladder}
\label{thermo}
We now study the thermoelectric property of this two-arm quasiperiodic ladder 
geometry. The quantitative measure of thermoelectric transport is manifested 
by the Seebeck coefficient which solely depends on the transmission characteristics
of the system under study showing the signature of the eigenspectrum. Experimental
measurement of the Seebeck coefficient~\cite{maci1,zotti,lunde,dubi} has been conducted at zero bias.
In that case, the system can be described by only a single Fermi level and the Seebeck
coefficient can be expressed in terms of $T(E)$ by means of the expression,
\begin{equation}
S = -\frac{\pi^2}{3} \frac{k_{B}^2 T}{|e|} \left( \frac{\partial \ln T(E)}{\partial E}
 \right)_{E_F}
\label{seebeck1}
\end{equation}
where $|e|$ is the charge of the electron, $k_B$ is the Boltzmann's constant and $T$ is
the mean temperature of the contacts. The above expression can be rewritten in terms
of Lorentz number as,
\begin{equation}
S = - |e| L_0 T \left( \frac{\partial \ln T(E)}{\partial E}
 \right)_{E_F}
\label{seebeck2}
\end{equation}
where $L_0 = \frac{\pi^2}{3} (\frac{k_{B}}{|e|})^2$ is called Lorentz number. 

In our case, the conductivity mechanism is described following the usual method
\begin{figure}[ht]
\centering
\includegraphics[clip,width=8.5cm,angle=0]{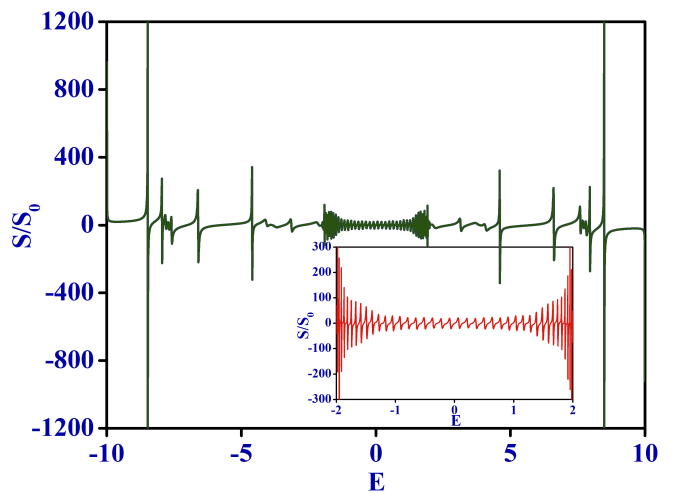}
\caption{(Color online) 
Plot of thermoelectric Seebeck co-efficient against energy(E). The term $S_0$ contains all the terms
of Eq.~\eqref{seebeck2}
excluding the derivative part. If the hopping integrals are taken in unit 
of $eV$ then $S$ has the unit of $\mu V K^{-1}$.}
\label{cuthermo}
\end{figure}
of multichannel transport~\cite{paramita1,paramita2,paramita3,santanu} (as mentioned earlier). Then by using the above expression for Seebeck coefficient we have obtained the behavior of thermoelectric transport of the 
quasiperiodic ladder network. Fig.~\ref{cuthermo} shows the variation of thermoelectric 
Seebeck coefficient of a finite size two-strand ladder arranged in a quasiperiodic 
coppermean geometry, sandwiched between pairs of semi-infinite ordered leads against the energy $E$ of the incoming electron. The plot shows unusual behavior due to the presence
of the minigaps in the electronic spectrum. From the figure it is clear that
the magnitude of the Seebeck coefficient diminishes slightly as we move from the 
outer regime to the central continuum part of energy spectrum. Interestingly, however,
besides small fluctuations near the subband edges, the Seebeck coefficient shows
strong oscillations near the minigaps. Moreover, throughout the whole range of energy
it shows several sharp transition between positive and negative magnitude mimicking a
\textit{Fano-like} nature. This repetitive sign change in the Seebeck coefficient 
as the Fermi energy varies has significant implications, indicating that the system may be tailored to exhibit n-or
 p-type properties using the same dopant (electron donors)
by carefully controlling the range of energy and also by selective choice of the 
tight-binding parameters of the network. It is to be noted that a similar behavior
of enhanced Seebeck coefficient due to miniband formation has also been reported 
by Balandin et. al.~\cite{balan} for three dimensional quantum dot array 
structures.

It is important to mention emphatically that the precise 
numerical correlation between
the parameters of the Hamiltonian can give a boost to engineer the band
spectrum at will. This immediately affects the transport and hence the thermoelectric
behavior. So with the aid of band tuning one can easily control the thermoelectric transport
in a subtle way by virtue of parametric correlations. This observation is quite
significant as this illustrates the advantage of tunable thermoelectricity of
such aperiodic geometry which could be convenient in the thermoelectric application
oriented aspects.

\section{Closing Remarks}
\label{conclu}
We have addressed the problem of observing transparent energy states in a system that is
quasi-one dimensional and non-translationally invariant. In the present arena of lithographic
and nanotechnology methodologies, and in particular, with the rapid advancement in the
controlled growth of optical lattices with trapped Bose-Einstein condensates, such systems can,
in principle, be made in practice. This has already achieved 
the milestone in perceiving the Anderson localization
of matter waves, and in the field of optics as well.
Our analysis reveals that, one can engineer sub-bands which are absolutely continuous and
holds Bloch functions only (even in the absence of translational invariance on a global scale).
The salient point of observation is that, for this one needs to correlate the numerical values
of a subset of the system parameters in a special way. This observation brings forth a highly
non-trivial variation of Anderson localization that can be monitored, and observed in low
dimensional systems.
The results are sound, specially when academic people are going for direct experimental
measurement of localization related properties in low dimensional lattices, and hence may
trigger further experiments on grafted lattices in low dimensions.

Specifically, here we have taken the example of copper mean sequence.
In this work we have computed the density of states (DOS) and transport
using a real space renormalization group method (RSRG).
A commutation relation between the potential and hopping matrix has been exploited
to work out the definite correlations between the parameters that can form 
absolutely continuous bands. The numerical correlations 
attained from the commutation play a pivotal role in the
\textit{crossover} of electronic eigen spectrum from
localization to completely continuum bands. This scenario obviously adds up a flavor of non-triviality in the
variation of the canonical case of Anderson localization. Finally,
 the 
tunable thermoelectric transport has been formulated for this quasiperiodic 
two-strand mesh.

\begin{acknowledgments}
A.M. gratefully acknowledges the financial support through
an INSPIRE fellowship [$IF 160437$] from DST, India. 
A.N. is thankful to the Department of Physics, University of Kalyani, W. B., India for
 providing the computational facility.
Both the authors are grateful for the stimulating
discussions regarding the results with Prof. Arunava Chakrabarti.
\end{acknowledgments} 

\end{document}